\documentclass[jkps,showpacs,showkeys,preprint]{revtex4}
\usepackage[dvips]{graphicx}
\usepackage{latexsym}

\begin{document}
\title{Statistical entropy and superradiance in 2+1 dimensional acoustic black holes}
 \author{Wontae \surname{Kim}}
   \email{wtkim@sogang.ac.kr}
 \author{Young-Jai \surname{Park}}
 \email{yjpark@sogang.ac.kr}
\author{Edwin J. \surname{Son}}
   \email{eddy@sogang.ac.kr}
 \affiliation{Department of Physics and Basic Science Research Institute,\\
  Sogang University, C.P.O. Box 1142, Seoul 100-611, Korea}
 \author{Myung Seok \surname{Yoon}}
   \email{yoonms@sejong.ac.kr}
 \affiliation{Department of Physics, Sejong University, Seoul, 143-747, Korea}


\begin{abstract}
We study ``draining bathtub'' as an acoustic analogue of a
three-dimensional rotating black hole. Rotating fluid near the sonic
horizon necessarily gives rise to the superradiant modes, which are
partially responsible for the thermodynamic quantities in this
rotating vortex-like hole. Using the improved brick-wall method, we
explicitly calculate the free energy of the system by treating the
superradiance carefully and obtain the desirable entropy formula.
\end{abstract}

\maketitle

\section{Introduction}
\label{sec:intro} As suggested by Bekenstein, a black hole may have
an intrinsic entropy proportional to the surface area at the event
horizon\cite{bekenstein}, and subsequently Hawking provided quantum
field theoretic calculations for the Schwarzschild black
hole\cite{hawking}. Since then, there has been much attention to the
statistical-mechanical origin of the entropy, especially for
rotating black holes\cite{mann,zt,ff,cho,hh}. In the brick-wall method,
quantum effects can be easily taken into account\cite{thooft}.
Introducing a brick-wall cutoff makes it possible to remove the
divergent term due to the infinite blue shift near the
horizon\cite{mukohyama,su,dlm}. The entropy from the brick-wall
method consists of mainly two parts: the most dominant term compared
to the logarithmic one gives the Bekenstein-Hawking entropy, and the
other represents a typical infrared contribution at large distances.
Although this original brick-wall method is useful for various
models\cite{thooft,mukohyama,su,dlm,gm,lk,hk,kkps,hkps,kksy,clp}, some
difficulties may arise because it is assumed that there exists a
thermal equilibrium between the black hole and the external field
even in a large spatial region. Obviously, this method cannot be
applied to a nonequilibrium system such as a system of
non-stationary space-time with two horizons because the two horizons
have different temperatures and the thermodynamical laws are also
invalid there. Solving these problems, a thin-film method as an
improved brick-wall method has been introduced\cite{hzk,zl}. In the
thin layer, local thermal equilibrium exists and the divergent term
due to large distance does not appear any more.

On the other hand, in Ref$.$\cite{unruh}, many black hole issues have
been treated as field theoretical problems in fluid because this
acoustic analogue was useful in that including its thermodynamics
such as Hawking radiation and entropy might be tested hopefully in
the laboratory. Moreover, a ``draining bathtub'' referred to as an
acoustic analogue of a rotating black hole had been well
defined\cite{visser,kko,vw,ls,bm,bcl}. However, the conventional
brick-wall cannot be applied in this model because it is impossible
that the angular velocities of particles have a same constant value
in whole region, while for the Ba\~nados-Teitelboim-Zanelli(BTZ)
black hole\cite{btz} the method was able to be used to examine some
results from superradiance effectively\cite{hk}. Thus, in this
paper, we would like to investigate the draining bathtub in terms of
the thin-film method, which is helpful since the angular speed near
the horizon can be approximately fixed to a constant. In
Sec.~\ref{sec:free}, the generic formulation of the free energy for
a rotating black hole is given in the grand canonical ensemble, and
it will be shown precisely why the thin-film method should be used
in our model. Then, in Sec.~\ref{sec:thermo}, the thermodynamic
quantities are calculated by treating superradiant and
nonsuperradiant(regular) modes more carefully. Finally, summary and
discussions are given in Sec.~\ref{sec:dis}.

\section{Formulation of the free energy for a rotating black hole}
\label{sec:free}

In order to calculate the entropy of a given system in the original
brick-wall method, we consider a quantum gas of scalar particles
confined within a box near the horizon of a black hole and introduce
a cut-off parameter as in Ref$.$\cite{thooft}. The free scalar field
is assumed to satisfy the Klein-Gordon equation given by
$(\Box+m^2)\psi = 0$ with boundary conditions
\begin{equation}
  \label{bc:period}
  \psi(r_H + h) = \psi(L) = 0,
\end{equation}
where $r_H$ is the horizon, and $r_H + h$ and $L$ represent the inner
and outer walls of a ``spherical'' box, respectively, and $h$ is an
infinitesimal cutoff parameter. Suppose that this system is in thermal
equilibrium at a temperature $T=\beta^{-1}$ with an external
reservoir. Then, in a stationary rotating axisymmetric black hole, a
partition $Z$ for an ideal bosonic gas in the grand canonical ensemble
is given by\cite{hk,wald}
\begin{equation}
  \label{Z}
  \ln Z = \sum_\lambda \ln\sum_{k=0}^\infty
  [e^{-\beta(\epsilon_\lambda - \Omega j_\lambda)}]^k,
\end{equation}
where $k=0,1,2,\cdots$ is an occupation number, $\epsilon_\lambda$ and
$j_\lambda$ denote the energy and angular momentum eigenvalues
for a single-particle state $\lambda$, respectively, and $\Omega$ is
the angular speed in equilibrium. The series in the partition function
(\ref{Z}) has a finite value for $\epsilon_\lambda - \Omega j_\lambda >0$, but
it becomes divergent for $\epsilon_\lambda - \Omega j_\lambda <0$, so
it is ill-defined. In order to resolve such problem caused by the
rotation of the geometry, we deal with the mode solutions of the
Klein-Gordon field carefully, which will be of the form,
$\psi(t,r,\phi) \sim e^{-i\omega t + i m\phi}$, for a rest observer at
infinity(ROI) because there exist two Killing vector fields denoted by
$\partial_t$ and $\partial_\phi$.

Note that the angular speed $\Omega$ in Eq.~(\ref{Z}) appears in the
thermodynamic first law for a reservoir, \textit{i.e.}, $dE = TdS +
\Omega dJ$ for a stationary rotating system. Besides, the angular
speed of a particle for a ROI should be restricted because no
particles can move faster than the speed of light. In fact, it takes a
value between the maximum $\Omega_+$ and the minimum $\Omega_-$ given by
\begin{equation}
  \label{ang:res}
  \Omega_\pm(r) = \Omega_0 (r) \pm \sqrt{(\partial_t \cdot
  \partial_\phi / \partial_\phi \cdot\partial_\phi)^2 - \partial_t
  \cdot \partial_t / \partial_\phi \cdot\partial_\phi},
\end{equation}
where $\Omega_0(r)$ is the angular speed of
Zero-Angular-Momentum-Observer(ZAMO)\cite{hk}. It is clear that both
$\Omega_\pm$ converge to the constant value of $\Omega_H \equiv
\Omega_0(r=r_H)$ near the horizon, so the angular speed of every
particles near the horizon can be always thought of as
$\Omega_H$. Since the dominant contribution to the physical quantities
of the system, such as total entropy, is attributed to the quantum gas
in the vicinity of the horizon, it is natural to assume that the
system is in equilibrium with a uniform angular speed $\Omega = \Omega_H$.

Before formulating a generic free energy for a rotating black hole,
the density function defined by $g(\omega,m) = \partial
n(\omega,m)/\partial \omega$ is introduced in order to calculate the
free energy strictly, where $n(\omega,m)$ is the number of mode
solutions whose frequencies, or energies, are all below $\omega$ for a
given value of angular momentum $m$. Thus, $g(\omega,m) d\omega$
represents the number of single-particle states whose energies lie
between $\omega$ and $\omega + d\omega$ and whose angular momenta are
$m$. Now, from the partition function~(\ref{Z}), the free energy $F$ is
obtained as
\begin{equation}
  \label{free:def}
  \beta F = - \ln Z = -\sum_m \int d\omega g(\omega,m) \ln \sum_k
  \left[ e^{-\beta(\omega - \Omega_H m)} \right]^k.
\end{equation}
It would be plausible to mention here that a ZAMO near the horizon($r
\approx r_H$) could measure only ingoing modes given by
$\psi_\mathrm{in}(x) \sim e^{-i\tilde\omega \tilde{t} + i \tilde{m}
  \tilde\phi}$, while a ROI would see both ingoing and outgoing ones,
where $\tilde{t} = t$, $\tilde\phi = \phi - \Omega_H t$, $\tilde
\omega = |\omega - \Omega_H m| >0 $, and $\tilde{m} = {\rm sgn}(\omega -
\Omega_H m) m$. Here, ${\rm sgn}(x)$ is 1 for $x > 0$ and $-1$ for $x <
0$. The ingoing wave near the horizon consists of two
parts; one is the so-called superradiant(SR) modes with $\omega -
\Omega_H m<0$, and the other is the nonsuperradiant(NS) modes with
$\omega - \Omega_H m >0$. Then, $e^{-i \tilde{\omega} t + i
\tilde{m} \tilde{\phi}} = e^{i \omega t - im\phi}$ for SR modes,
and $e^{-i \tilde{\omega} t + i \tilde{m} \tilde{\phi}} = e^{-i
\omega t + im\phi}$ for NS modes. Since only the ingoing modes are
considered near the horizon, $(\epsilon,j)$ has the value of
$(\omega,m)$ for single-particle states with the NS modes, while
$(\epsilon,j)$ becomes $(-\omega,-m)$ for the SR ones. Separating the
SR modes from the NS ones, the free energy~(\ref{free:def}) should be
replaced by $F = F_\mathrm{NS} + F_\mathrm{SR}$ with
\begin{eqnarray}
  & & \beta F_\mathrm{NS} = \sum_{m} \int_{\omega > \Omega_H m}
      d\omega g(\omega,m) \ln \left[ 1 - e^{-\beta(\omega - \Omega_H
      m)} \right], \label{F:NS} \ \\
  & & \beta F_\mathrm{SR} = \sum_{m} \int_{\omega < \Omega_H m}
      d\omega g(\omega,m) \ln \left[ 1 - e^{\beta(\omega - \Omega_H
      m)} \right]. \label{F:SR} \
\end{eqnarray}
Note that $\omega$ is positive definite, and the density
functions are given by $g(\omega,m) = \partial n(\omega,m) / \partial
\omega$ for the NS modes and $g(\omega,m) = - \partial n(\omega,m) / \partial
\omega$ for the SR ones. Both Eqs.~(\ref{F:NS}) and
(\ref{F:SR}) can be obtained from $\beta F = - \sum_{\tilde{m}} \int
d\tilde{\omega} g(\tilde{\omega},\tilde{m}) \ln \sum_k \exp(-k \beta
\tilde{\omega})$, where $g(\tilde{\omega},\tilde{m}) = \partial
n/\partial \tilde{\omega}$.

On the other hand, the angular speed of any particle cannot reach
$\Omega_H$ over a critical radius in our model because of the
restriction for $\Omega$ from Eq.~(\ref{ang:res}), which will be explicitly
shown in the following section so that global thermal equilibrium
cannot be achieved. Instead, if we consider scalar particles
confined within a thin layer near the horizon\cite{hzk,zl}, their
angular speeds naturally take the same value of a constant $\Omega_H$
due to local thermal equilibrium, and the thin-layer method is useful
to find the thermodynamic quantities in our model.

Apparently, the degrees of freedom of a field within a thin layer
near the horizon play a major role in the calculation of the entropy
of a black hole; hence global thermal equilibrium is not necessary
any more because particles are assumed to be distributed only in the
narrow region. Since it is well known that Hawking radiation is
derived from the vacuum fluctuation near the horizon, the
Bekenstein-Hawking entropy should be associated with the field in
this narrow region, where thermal equilibrium exists locally and
statistical mechanics remains valid. This local thermal equilibrium
is the main postulate of thin-film method, and the thermodynamic
properties such as pressure and temperature near the horizon is
assumed to vary slightly. The thickness of layer is supposed to be
so small on a macroscopic scale that the physical quantities can be
considered to be constant and that the narrow region can be locally
in thermal equilibrium. Also, it is supposed to be very large on a
microscopic scale to make sure that statistical mechanics remains
valid. Specifically, in our model, the outer boundary $L$ of
spherical box in Eq.~(\ref{bc:period}) is replaced by $r_H + h +
\delta$, where the parameter $\delta$ is a positive physical small
quantity related to the thickness of the layer. It means that
$\delta$ has the scale over Plank length, but the brick-wall cutoff
$h$ is very small compared to the Plank length.

\section{Thermodynamic quantities}
\label{sec:thermo}

We now set up an acoustic analogue of a rotating BTZ
black hole in order to investigate its thermodynamics with
superradiance taken into account.
In the irrotational fluid, the propagation of sound waves is governed
by an equation of motion\cite{unruh},
\begin{equation}
  \label{eom}
  \Box \psi = \frac{1}{\sqrt{-g}} \partial_\mu (\sqrt{-g} g^{\mu\nu}
  \partial_\nu \psi)=0,
\end{equation}
where $\psi$ is the fluctuation of the velocity potential
interpreted as a sonic wave function, and the metric is given by
\begin{equation}
  \label{metric:general}
  g_{\mu\nu} = \frac{\rho_0}{c} \left(
    \begin{array}{cc}
      -(c^2 - v_0^2) & - v_0^i \\
      -v_0^j & \delta_{ij}
    \end{array}\right) \qquad \mathrm{with}\ i,j = 1,2,3,
\end{equation}
where $c$ is the speed of sound wave, $\rho_0$ and $v_0^i$ are the
mass density and the velocity of the mean flow, respectively,
$\delta_{ij}$ is the Kronecker delta, and $v_0^2 = \delta_{ij} v_0^i
v_0^j$. Note that the velocity potential is linearized as $\Psi =
\psi_0 + \psi$, and $\vec{v}_0 = \vec\nabla \psi_0$.

We then consider a draining bathtub fluid flow described as a
$(2+1)$-dimensional flow with a sink at the origin. If the metric is
stationary and axisymmetric, the equation of continuity, Stokes'
theorem, and conservation of angular momentum yield that $\rho_0$ is
constant and $\psi_0 (r,\phi) = - A \ln (r/a) + B\phi$, where $a$,
$A$, and $B$ are arbitrary real positive constants\cite{visser}.
Then, the velocity of the mean flow is given by $\vec{v}_0 = -
\hat{r} (A/r) + \hat{\phi} (B/r)$.

Now, let us consider the draining vortex case with $A \neq 0$.
Dropping a position-independent prefactor from the
metric~(\ref{metric:general}), the acoustic line element for the
draining bathtub is obtained as
\begin{equation}
  \label{line}
  ds^2 = -c^2 dt^2 + \left( dr + \frac{A}{r}dt \right)^2 + \left(r
  d\phi - \frac{B}{r}dt \right)^2,
\end{equation}
where the radii of the horizon and the ergosphere are
\begin{equation}
  \label{horizon}
  r_H = \frac{A}{c}, \qquad r_e = \frac{\sqrt{A^2 + B^2}}{c},
\end{equation}
respectively. However, the metric~(\ref{line}) makes it difficult to
calculate thermodynamic quantities because of its ($t$,$r$)-component.
Fortunately, this can be overcome by a coordinate transformation in the
exterior region of $A/c < r < \infty$. Using the transformation given
by\cite{bm,bcl}
\begin{equation}
  \label{transf}
  dt \rightarrow dt + \frac{Ar}{r^2c^2 - A^2} dr, \qquad d\phi \rightarrow
  d\phi + \frac{AB}{r(r^2c^2 - A^2)} dr,
\end{equation}
the metric~(\ref{line}) can be rewritten as the conventional form,
\begin{equation}
  \label{metric}
  ds^2 = - N^2 dt^2 + N^{-2} dr^2 + r^2 (d\phi - \Omega_0 dt)^2
\end{equation}
with
\begin{equation}
  \label{func}
  N^2(r) = 1 - \frac{A^2}{c^2r^2} = \frac{r^2 - r_H^2}{r^2}, \qquad
  \Omega_0(r) = \frac{B}{cr^2} = \Omega_H \frac{r_H^2}{r^2},
\end{equation}
where we rescaled time coordinate by $c$ for simplicity, and
$\Omega_H = B/(cr_H^2)$. Note that the metric~(\ref{metric}) is
similar to that of a rotating BTZ black hole, but two metrics have a
little difference: although setting $J = 2B/c$ gives the same angular
speed $\Omega_0 (r)$ of ZAMO, the lapse function $N(r)$ has a
different form from that of BTZ black hole, which is explicitly given
by $N^2 = (r^2 - r_+^2)(r^2 - r_-^2)/(r^2l^2)$.

\begin{figure}[t]
\includegraphics[width=0.6\textwidth]{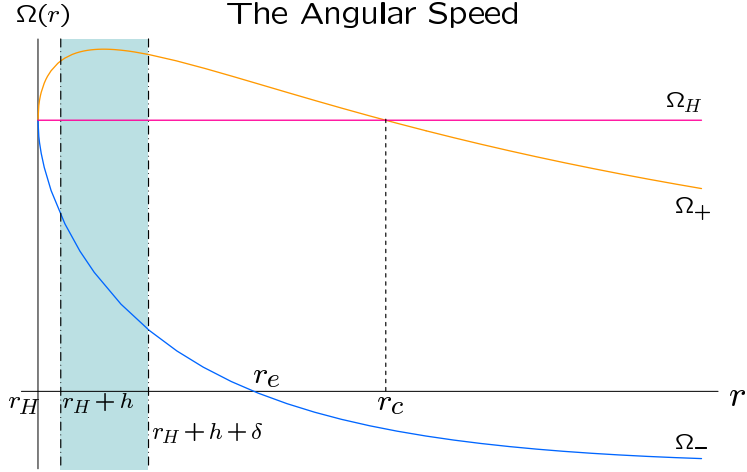}
\caption{The possible angular speed of a particle lies between the
  upper and lower curves, which denote the maximum and minimum
  angular speeds with respect to $r$, respectively. The shaded
  area represents the thin layer, which is located inside the
  ergoregion.}
\label{fig:ang}
\end{figure}

Now, the maximum and minimum angular speeds are obtained from
Eqs.~(\ref{ang:res}) and (\ref{metric}) as
\begin{equation}
  \label{omega+-}
  \Omega_\pm(r) = \Omega_0(r) \pm \frac{N(r)}{r}.
\end{equation}
As mentioned before, the angular speed of every particle is
$\Omega_H$ in the vicinity of the horizon since $\Omega_\pm$ goes to
$\Omega_H$ as $r \rightarrow r_H$. Note that there exists a critical
radius $r_c = \sqrt{r_H^2 + \Omega_H^{-2}}$, where the maximum angular
speed $\Omega_+$ is equal to $\Omega_H$. This means that no particle can move
along with the angular speed of $\Omega_H$ over $r_c$ as shown in
Fig.~\ref{fig:ang}, so the spherical box should be located inside the
critical radius $r_c$. Moreover, as discussed in the previous section,
the radius of its outer boundary should be smaller than that of the
ergosphere. Then, the radial part of the sonic wave
satisfies
\begin{equation}
  \label{eom:r}
  rN^2 \frac{d}{dr} \left[ rN^2 \frac{d}{dr}\psi_{\omega m}(r) \right]
  + r^2 N^4 k^2 (r;\omega,m) \psi_{\omega m}(r) = 0,
\end{equation}
where $k^2(r;\omega,m) = N^{-4}(\omega - \Omega_+ m)(\omega - \Omega_-
m)$. It can be easily shown that the function $k(r;\omega,m)$ plays
the role of the momentum eigenvalue in the WKB
approximation. Therefore, in the thin layer with the range of $r_H + h
< r < r_H + h + \delta$, the discrete energy $\omega$ is related to
the total number $n(\omega,m)$ by
\begin{equation}
  \label{number}
  \pi n(\omega,m) = \int_{r_H + h}^{r_H + h + \delta} dr \, k(r;\omega,m),
\end{equation}
where $k(r;\omega,m)$ is set to be zero if $k^2(r;\omega,m)$ becomes
negative for given $(\omega,m)$\cite{thooft}. The contribution of
$k$ to the following calculations is dominant near the horizon
because $k$ is approximately $N^{-2}$ and diverges as $r$ goes to
$r_H$. This tells us that the thin-film method is valid.

Then, we now evaluate the free energy of total system. The free energy for
NS modes~(\ref{F:NS}) is written as
\begin{eqnarray}
  \beta F_\mathrm{NS}\!\! &=\!\!& \sum_{m} \int_{\omega>\Omega_H m}
        d\omega \frac{\partial}{\partial\omega}\left[ \frac{1}{\pi}
        \int_{r_H + h}^{r_H + h + \delta} dr \, k(r;\omega,m) \right]
        \ln \left[ 1- e^{-\beta(\omega - \Omega_H m)} \right] \
        \nonumber \\
     &=\!\!& -\frac{\beta}{\pi} \int_{r_H + h}^{r_H + h + \delta} dr \sum_m
        \int d\omega \frac{k(r;\omega,m)}{e^{\beta(\omega - \Omega_H m)} -
        1} \ \nonumber \\
     & & + \frac{1}{\pi} \int_{r_H + h}^{r_H + h + \delta} dr \sum_m
        k(r;\omega,m) \ln \left.\left[ 1- e^{-\beta(\omega - \Omega_H m)}
        \right]\right|_{\omega_\mathrm{min}(m)}^{\omega_\mathrm{max}(m)},
        \ \label{free:NS}
\end{eqnarray}
by using the integration by parts with respect to $\omega$. For the
sake of convenience, the free energy $F_\mathrm{NS}$ for NS modes is
divided into two parts, which describe states with positive and
negative angular momentum, \textit{i.e.}, $F_\mathrm{NS} =
F_\mathrm{NS}^{(m>0)} + F_\mathrm{NS}^{(m<0)}$, where
\begin{eqnarray}
  F_\mathrm{NS}^{(m>0)} \!\!\! &=\!\!\!& -\frac{1}{\pi} \int_{r_H +
      h}^{r_H + h + \delta} dr N^{-2} \int_0^\infty dm \int_{\Omega_+
      m}^\infty d\omega \frac{\sqrt{(\omega-\Omega_+ m)(\omega - \Omega_-
      m)}}{e^{\beta(\omega - \Omega_H m)}- 1}, \label{free:NS+}\\
  F_\mathrm{NS}^{(m<0)} \!\!\! &=\!\!\!& -\frac{1}{\pi} \int_{r_H +
      h}^{r_H + h + \delta} dr N^{-2} \int_{-\infty}^0 dm \int_0^\infty
      d\omega \frac{\sqrt{(\omega-\Omega_+ m)(\omega - \Omega_-
      m)}}{e^{\beta(\omega - \Omega_H m)}- 1} \nonumber\\
    & & - \frac{1}{\pi\beta} \int_{r_H + h}^{r_H + h + \delta} dr N^{-2}
      \int_{-\infty}^0 dm \sqrt{\Omega_+\Omega_- m^2} \ln(1-
      e^{\beta\Omega_H m}). \label{free:NS-}
\end{eqnarray}
Similarly, the free energy for SR modes~(\ref{F:SR}) becomes
\begin{eqnarray}
  F_\mathrm{SR} \!\! &=\!\!& - \frac{1}{\pi} \int_{r_H + h}^{r_H + h +
    \delta} dr N^{-2} \int_0^\infty dm \int_0^{\Omega_- m} d\omega
    \frac{\sqrt{(\omega-\Omega_+ m)(\omega - \Omega_-
    m)}}{e^{-\beta(\omega - \Omega_H m)}- 1} \nonumber \\
  & & + \frac{1}{\pi\beta} \int_{r_H + h}^{r_H + h +
    \delta} dr N^{-2} \int_0^\infty dm \sqrt{\Omega_+\Omega_- m^2} \ln(1-
      e^{-\beta\Omega_H m}). \label{free:SR}
\end{eqnarray}
Fortunately, the second terms of Eqs.~(\ref{free:NS-}) and
(\ref{free:SR}) are exactly cancelled, but large portion of tedious
calculations is still required to be evaluated.
After evaluating the above integrations with respect to $\omega$ and
$m$, the expression of total free energy is obtained as
\begin{equation}
  \label{free:exact}
  F = - \frac{\zeta(3)}{4\beta^3} \int_{r_H + h}^{r_H + h + \delta}
  dr \frac{(\Omega_+ - \Omega_-)^2}{N^2(\Omega_+ - \Omega_H)^\frac32
  (\Omega_H - \Omega_-)^\frac32}.
\end{equation}
Then, substituting Eqs.~(\ref{func}) and (\ref{omega+-}) into
Eq.~(\ref{free:exact}), the total free energy of our system is reduced
to
\begin{equation}
  \label{free:cal}
  F = - \frac{\zeta(3) r_H^2}{\beta^3} \left[ \sqrt{\frac{r_H}{2h}} -
  \sqrt{\frac{r_H}{2(h+\delta)}}  +
  O(\sqrt{h},\sqrt{h+\delta}) \right]
\end{equation}
in the leading order. Note that there are no logarithmically divergent
terms in the total free energy~(\ref{free:cal}) because those are
remarkably cancelled as well as in the rotating BTZ black hole
case\cite{hk}. In addition, there exists no infrared divergence since
the large distance is out of consideration, while infrared divergent
terms remaining in total free energy were cut off from consideration
in Refs$.$\cite{thooft,hk}. It seems appropriate to comment here that
for the limiting case of non-rotating acoustic black hole, $B=0$,
there are no SR modes due to the fact that the angular speed of
horizon vanishes, $\Omega_H=0$; in addition, there are no critical
radius $r_c$ and no ergoregion due to $r_e = r_H$. Thus, one might think in
this case having only NS modes the above result should be recast and
give different value; however, the very same result is obtained so
that Eq.~(\ref{free:cal}) is remained valid in the limit of
$\Omega_H\to0$. And since SR modes are distinguished from NS ones
for ROI and not for ZAMO near the horizon, it is reasonable that the
result from considering the superradiant modes in the rotating
acoustic black hole is the same as the non-rotating one.

On the other hand, the surface gravity is given by $\kappa_H^2 \equiv
-\left.\frac12\nabla^\mu\chi^\nu \nabla_\mu\chi_\nu \right|_{r=r_H}
= 1/r_H^2$, where we used an appropriate Killing field near the
horizon, $\chi^\mu = (\partial_t + \Omega_H
\partial_\phi)^\mu$\cite{wald:book}, and then the temperature becomes
\begin{equation}
  \label{temp}
  T_H = \beta_H^{-1} = \frac{\kappa_H}{2\pi} = \frac{1}{2\pi r_H}.
\end{equation}
Using the thermodynamic relation $S=\beta^2\partial F/\partial\beta
|_{\beta=\beta_H} = -3 \beta F|_{\beta=\beta_H}$, the entropy of this
system is obtained from the free energy~(\ref{free:cal}) as
\begin{equation}
  \label{entropy}
  S = \frac{3\zeta(3)}{4\pi^2} \left[ \sqrt{\frac{r_H}{2h}} -
  \sqrt{\frac{r_H}{2(h+\delta)}}  +
  O(\sqrt{h},\sqrt{h+\delta}) \right],
\end{equation}
and it can be rewritten in terms of the thin-wall cutoffs as
\begin{equation}
  \label{S:cutoff}
  S = \frac{3\zeta(3)}{8\pi^3}
  \frac{\bar\delta}{\bar{h}(\bar{h} + \bar\delta)}\ell +
  O(\bar{h},\bar{h} + \bar\delta),
\end{equation}
where the cutoffs were defined as $\bar{h} \equiv \int_{r_H}^{r_H + h}
\sqrt{g_{rr}} \, dr \approx \sqrt{2 r_H h}$ and $\bar{\delta} \equiv
\int_{r_H + h}^{r_H + h + \delta} \sqrt{g_{rr}}\, dr \approx \sqrt{2
  r_H (h + \delta)} - \sqrt{2 r_H h}$ in the leading order, and
$\ell \equiv \left. \int_{0}^{2\pi} \sqrt{g_{\phi\phi}}\, d\phi
\right|_{r=r_H} = 2\pi r_H$ is the circumference of the horizon. Note
that $\bar{h}$ is called the brick-wall cutoff, and $\bar{\delta}$
represents the thickness of the thin layer. Then,
setting $\bar{\delta} / [\bar{h} ( \bar{h} + \bar{\delta} )] = 16
\pi^3 / [3 \zeta(3) \ell_p]$, or equivalently $\bar{h} =
(\bar{\delta}/2) [ -1 + \sqrt{1 + 3 \zeta(3) \ell_p / (4 \pi^3
  \bar{\delta})}]$, the entropy of sonic wave becomes finite and
equivalent to the Bekenstein-Hawking entropy in the leading order,
\begin{equation}
  \label{S:btz}
  S = \frac{4\pi r_H}{\ell_p} = S_\mathrm{BH},
\end{equation}
where the three-dimensional Plank length is chosen as $\ell_p \equiv
\hbar G / c^3$. It is plausible to make sure that the brick-wall
cutoff becomes a universal value of $\bar{h} = [3\zeta(3)/(16\pi^3)]
\ell_p$ in the leading order if $\bar\delta$ is larger than $\ell_p$,
which is set to be one in the following calculations.

Next, let us calculate the other thermodynamic quantities, such as
the angular momentum of a matter particularly interpreted as a
phonon in our model,
\begin{equation}
  \label{J}
  J_\mathrm{matter} = - \left. \frac{\partial F}{\partial \Omega_H}
    \right|_{\beta=\beta_H}
   = \frac{3\zeta(3)r_H^2\Omega_H}{8\pi^3}
  \frac{\bar\delta}{\bar{h}(\bar{h} + \bar\delta)} ,
\end{equation}
where the partial derivative was evaluated from
Eq.~(\ref{free:exact}). Note that substituting the expression of
cutoff $\bar h$ into Eq.~(\ref{J}), the angular momentum is given by
\begin{equation}
  \label{J:thick}
  J_\mathrm{matter} = 2 \Omega_H r_H^2 = \frac{2B}{c}.
\end{equation}
The internal energy of the system in the frame of a ROI is explicitly
calculated as
\begin{equation}
  \label{E}
  E = F_H + \beta_H^{-1} S + \Omega_H J_\mathrm{matter} = \frac43 +
  2r_H^2\Omega_H^2 = \frac43 + \frac{2B^2}{A^2},
\end{equation}
where $F_H = F|_{\beta=\beta_H}$. In the limit of non-rotating
acoustic black holes of $J_\mathrm{matter} \rightarrow 0$, or
equivalently $B \to 0$, it can be easily seen that the total energy
$E$ has the minimum value of $4/3$.

Finally, it seems to be appropriate to comment on the perfect vortex
case. If we take the limit of pure spinning acoustic black holes of
$A \rightarrow 0$, the internal energy in Eq. (29) is undefined.
Therefore, we must independently analyse this case whose spacetime
represents a fluid with a non-radial flow. But, there does not exist
the event horizon any more in this case. Therefore, it is
meaningless to consider the analogy between gravitational and
acoustic black holes.

\section{Discussion}
\label{sec:dis} In this paper, we have studied the thermodynamics of
a rotating acoustic black hole involving the superradiant modes for the
draining vortex case($A \neq 0$) as an acoustic analogue of a black
hole in three dimensions. In order to overcome some difficulties in
applying the original brick-wall method to our model, the thin-film
method has been introduced as an improved brick-wall one. And using
this method we have obtained the thermodynamic quantities such as
free energy, entropy, angular momentum, and internal energy of the
thin-layer under thermal equilibrium with the black hole. The
definition of thermodynamic black hole entropy was chosen in
Eq.~(\ref{S:btz}) as $S_{BH} = 2\ell/\ell_p$ following that of BTZ
black hole\cite{btz} to fix the brick-wall cutoff $\bar{h}$, where
the leading order of the entropy becomes $S \approx S_{BH}(1 -
\bar{h}/\bar{\delta})$ for a universal value of brick-wall cutoff.
Recovering the physical dimension, the angular
momentum~(\ref{J:thick}) and the internal energy~(\ref{E}) becomes
$J_\mathrm{matter}=2c^2B/G$ and $E/c^2 = (4/3+2B^2/A^2)c^2/G$,
respectively.

As for the case of the limit of $A \rightarrow 0$, the internal
energy~(\ref{E}) diverges. In fact, the metric~(\ref{metric}) of
$A=0$ seems to describe a pure rotation without horizons. However,
this limit could not be taken since it has a naked singularity at
$r=0$. Therefore, we should consider a different form from the
metric~(\ref{metric}) in order to deal with the pure rotation. Also,
in the pure rotation, the brick-wall method could not be used to
calculate thermodynamic quantities since the particles are
distributed in whole region and it is impossible for the particles
to fix the angular velocities to a special value.

Finally, for the purpose of checking the stability of the system,
the heat capacity can be calculated as\cite{bcm}
\begin{equation}
  \label{C_J}
  C_J \equiv \left. \left( \frac{\partial E}{\partial T} \right)_J
     \right|_{\beta=\beta_H}
  = 2S,
\end{equation}
where we used the first law of thermodynamics, $dE = TdS + \Omega_H dJ$,
and the thermodynamic relation between the entropy and the free energy
from Eq.~(\ref{free:exact}). Since the
entropy is always positive, the heat capacity~(\ref{C_J}) is positive,
and this means that the rotating acoustic black hole is thermodynamically
stable. And also it can be easily shown that the curvature scalar of
background geometry is positive everywhere as $R = 2[r_H^2 +
(J/2)^2]/r^4$.


\begin{acknowledgments}
We would like to thank G. Kang for useful discussions. W. Kim and
E. J. Son are supported by the Science Research Center Program of the
Korea Science and Engineering Foundation through the Center for
Quantum Spacetime\textbf{(CQUeST)} of Sogang University with grant
number R11 - 2005 - 021. Y.-J. Park is supported by the Korea Research
Foundation Grant funded by the Korean Government
(KRF-2005-015-C00105). M. S. Yoon is supported by the Korea Research
Foundation Grant funded by the Korean Government
(MOEHRD)(KRF-2005-037-C00017).
\end{acknowledgments}


\end{document}